\documentclass[12pt]{revtex4}
 \begin{document}
 
 \centerline{\bf \Large Stephen Hawking (1942-2018) }
 \centerline{{\bf \large Towards a Complete Understanding of the Universe}  \footnote{A slightly edited version of of an article solicited by the {\sl Proceedings of the (US) National Academy of Sciences,} for their {\sl Retrospectives} section.  It appeared in the issue of May 7, 2018, 201806196,
 https://doi.org/10.1073/pnas.1806196115}}
 
\centerline{ \large \bf  James Hartle}
\centerline{ \large University of California, Santa Barbara and Santa Fe Institute}

 \vskip .25in
 \large

 Stephen Hawking overcame the limitations  of a debilitating disease to make major contributions to science. He did this through remarkable persistence, determination, conviction, courage, and will.  He was supported in this by his  family, his students, and his colleagues.  

Stephen said  ``My goal is simple. It is a complete understanding of the universe.''  Stephen's many contributions to science moved us significantly toward that goal. Much of his work was  in two major areas --- cosmology and the physics of black holes. I will only  mention one  contribution in each that stands out  ---  his work on the beginning of the universe and  his work on the end of black holes \footnote{This article is a length limited informal description of  the two of Stephen's major achievements in science for which he is perhaps best known.  For more on the author's personal experiences with Stephen see ``Working with Stephen'', arXiv:1711.09071 and  the authorÕ's contribution to the March 14 article in {\sl Physics Today} in which Stephen Hawking is remembered
by his colleagues. https://goo.gl/dRgf3q), arXiv:1803.09197}.

 {\it The Beginning of the Universe:}  
 
  Stephen's Ph.D. advisor Dennis Sciama started him working on cosmology.  An important question in the '60s was: ``Did the universe have a beginning?''
Observations and the Einstein equation suggested that the early universe was very hot. But did the universe have a beginning or was it first big and then contracted, heated,  and then re-expanded? The singularity theorems worked out by Stephen and Roger Penrose  settled the matter at the level of classical gravitational physics.  The universe did have a beginning in a big bang of infinite temperature and density before which we can not not see,  and at which the classical Einstein equation breaks down. Much of our confidence that there was a big bang rests on these theorems.

Stephen's  singularity theorems meant that the universe could not begin with the three space and one time dimensions of classical physics. But it could start with four space dimensions that later made a {\it quantum} transition to three dimensions of space and one of time. 

A quantum universe is described by a wave function representing its quantum state.  Stephen and I provided one for our universe  --- the no-boundary wave function. Its semiclassical approximation does describe a transition from four space dimensions to three space and one time. Stephen provided a new birth for the field {\it quantum} cosmology.

{\it The End of Black Holes:} 

When a massive star runs out of  thermonuclear fuel to burn to get the heat  to support itself against the force of gravity it can collapse and make a black hole --- a region of spacetime where,  according to classical physics, the gravitational attraction is so strong that nothing inside, not even light,  can get out.

Stephen's work on black holes began about 1970 when in a `eureka moment' he realized that the powerful global techniques used for the singularity theorems could be applied to understand the properties of black holes in new levels of generality. What do we mean generally by a black hole? What is the most general black hole allowed by general relativity?  Stephen contributed essential pieces to the answers of these questions. 

In 1974,  Stephen revolutionized the field of black hole physics by showing that black holes are not black when quantum mechanics is taken into account. They have a temperature and radiate (the Hawking radiation).   As he put it ``black holes can be white hot''. This remarkable result touches on the most fundamental issues in physics because it means that black holes evaporate and disappear. More than 40 years  later many scientists are still working to understand the implications of this for fundamental physics.

{\it The Unique Stephen:} 

 Stephen had a number of qualities that helped him achieve what he did.

There is the the unity of his physics.  Black holes and the beginning of the universe have several similarities. First, the key features of both are accessible by the global techniques that Stephen and others developed. Second, classically they both exhibit singularities --- the big bang in the past of our universe, and the inevitable future singularity inside black holes hidden from us by its event horizon.

 Stephen 
 had a remarkable ability to  focus on the essentials of a problem and ignore the clutter. He was helped in this by a commitment to making predictions for observation. Stephen always knew the right question to ask whose answer would move us forward. 

But he also knew the right thing to  give up. Advances in physics of course involve new ideas. But many times  they also involve giving up on cherished current ideas. Stephen taught us that we have to give up on the idea that black holes were black to get to the Hawking radiation.  We have to give up on the idea that spacetime always had three space dimensions and one time dimension to get to a quantum theory of the big bang.. At the time these ideas were surprising, but Stephen loved to surprise. Today many of them seem simple, natural, and even inevitable, but that was his  genius.

{\it Stephen and Us:}
Stephen's scientific legacy does not lie only  in his papers. It lies also  in the inspiration  he provided those of us that followed him.  That is most concretely exhibited by the graduate students he supervised many of whom have become distinguished scientists themselves. It lies also in his outstanding ability to communicate science to the general public in books like his best selling `Brief History of Time' \footnote{S. Hawking, {\it A Brief History of Time}, {Bantam Press, London, 1988.}} and in his  beautifully crafted public lectures.  Finally, it lies also in the example that he set with his life which has been an inspiration to many.

 \end{document}